\documentclass[journal]{IEEEtran}
\usepackage{graphicx}
\graphicspath{ {./images/} }
\usepackage{caption}
\usepackage{amsmath}

\ifCLASSINFOpdf
\else
\fi
\usepackage{xcolor}
\IEEEoverridecommandlockouts \IEEEpubid{\makebox[\columnwidth]{Paper accepted for presentation at IEEE LATINCOM 2020 \hfill}
\hspace{\columnsep}\makebox[\columnwidth]{ }}
\begin{document}
%
\title{Review: Deep Learning Methods for Cybersecurity and Intrusion Detection Systems}

%
%
%

\author{Mayra Macas, Chunming Wu\\~\IEEEmembership{College of Computer Science and Technology, Zhejiang University\\
No. 38 Zheda Road, Hangzhou 310027, China\\Emails:\{mayramacas11, wuchunming\}@zju.edu.cn}

\thanks{Mayra Macas is supported by the Chinese government scholarship CSC Reg. No.: 2017GBJ005834}
}

\maketitle

\begin{abstract}
As the number of cyber-attacks is increasing, cybersecurity is evolving to a key concern for any business. Artificial Intelligence (AI) and Machine Learning (ML) (in particular Deep Learning - DL) can be leveraged as key enabling technologies for cyber-defense, since they can contribute in threat detection and can even provide recommended actions to cyber analysts. A partnership of industry, academia, and government on a global scale is necessary in order to advance the adoption of AI/ML to cybersecurity and create efficient cyber defense systems. In this paper, we are concerned with the investigation of the various deep learning techniques employed for network intrusion detection and we introduce a DL framework for cybersecurity applications.
\end{abstract}

%
\IEEEpeerreviewmaketitle

\section{Introduction}
\label{sec:intro}
By 2023, it is anticipated that the number of IP-connected devices will be three times larger than the global population, producing up to 4.8 ZB of IP traffic annually, as pointed out by Cisco \cite{IEEEhowto:23}. This accelerated increase raises overwhelming security challenges. Thus, how to identify network attacks is a key problem that should never been overlooked. As an important and active security mechanism, Intrusion Detection has become the key technology of network security. The objective of the Intrusion Detection Systems (IDSs) is to identify unusual access or attacks on internal network security, which can be implemented on a misuse or anomaly basis~\cite{IEEEhowto:2}. 

Misuse-based techniques search for specific patterns or signatures of attacks in system calls, network traffic, and so forth. However, these techniques are not useful against zero-day attacks because they require regular updates to databases that store rules and signatures. Anomaly-based techniques, on the other hand, identify abnormalities by distinguishing it from normal behavior (i.e., they search for significant differences from normal traffic). Such techniques can detect unknown attacks and cannot be easily avoided by the attackers because normal activity is tailored to each particular user, application, and network. However, their drawbacks include the production of high FPRs, since any previously unseen traffic or system is flagged as a potential malicious attack, and the need of performing individual training for every deployment.

Within this context, the scientific community has studied and designed intrusion detection systems based on DL. DL models learn representations of raw input with multiple levels of abstractions with little manual intervention. The final product of complex layer-based processing is a set of high-level features. The feature representation in the higher layers of a DL model principally designed for the classification task increases the aspects of the input data that are essential for the discrimination, while ignoring irrelevant details. As a consequence, the emergence of a new variant of cyberattack (e.g., zero-day malware) does not require extracting features from scratch. Instead, the features that increase the generalization of the classification task will be automatically engineered by a vast number of non-linear layers. Therefore, complex cybersecurity problems can be solved efficiently by carefully using and customizing DL models.

The remainder of this paper is organized as follows. Section~\ref{sec:Background} provides an introduction to DL along with an overview of its state-of-the-art models, while in Section~\ref{sec:framework} a DL framework for cybersecurity applications is introduced. Finally, Section~\ref{sec:review} constitutes a literature survey of the most representative DL-based systems employed for network intrusion detection, matching the proposed methodologies with the dataset(s) used for performance evaluation. In Section \ref{sec:Conclusion} we conclude the paper.
\section{Background}
\label{sec:Background}
\subsection{Deep Learning}
DL is a set of algorithms that are based on Artificial Neural Networks (ANNs). These computational models are more or less designed in analogy with biological brain modules. Nowadays ANNs are back in fashion under the term DL. One of the milestones in this story occurred in 2006, when Geoffrey Hinton's lab was able to train efficiently a deep network for reconstructing high-dimensional input vectors from a small central layer~\cite{IEEEhowto:16}. After that, this technology has been adopted in various fields of AI, such as image retrieval, image recognition, natural language processing, search engines, and information retrieval. One advantage of DL architectures compared to traditional ANNs is learning hidden features from raw data \cite{IEEEhowto:19}. The fact that each layer trains on a set of features based on the outputs of the previous layers creates a hierarchy of features. According to this, the inner-most layers are able to recognize more complex features, because they aggregate and recombine features from the previous layers.
      
\begin{figure}[h]
\centering
\captionsetup{justification=centering}
\includegraphics[width=8cm]{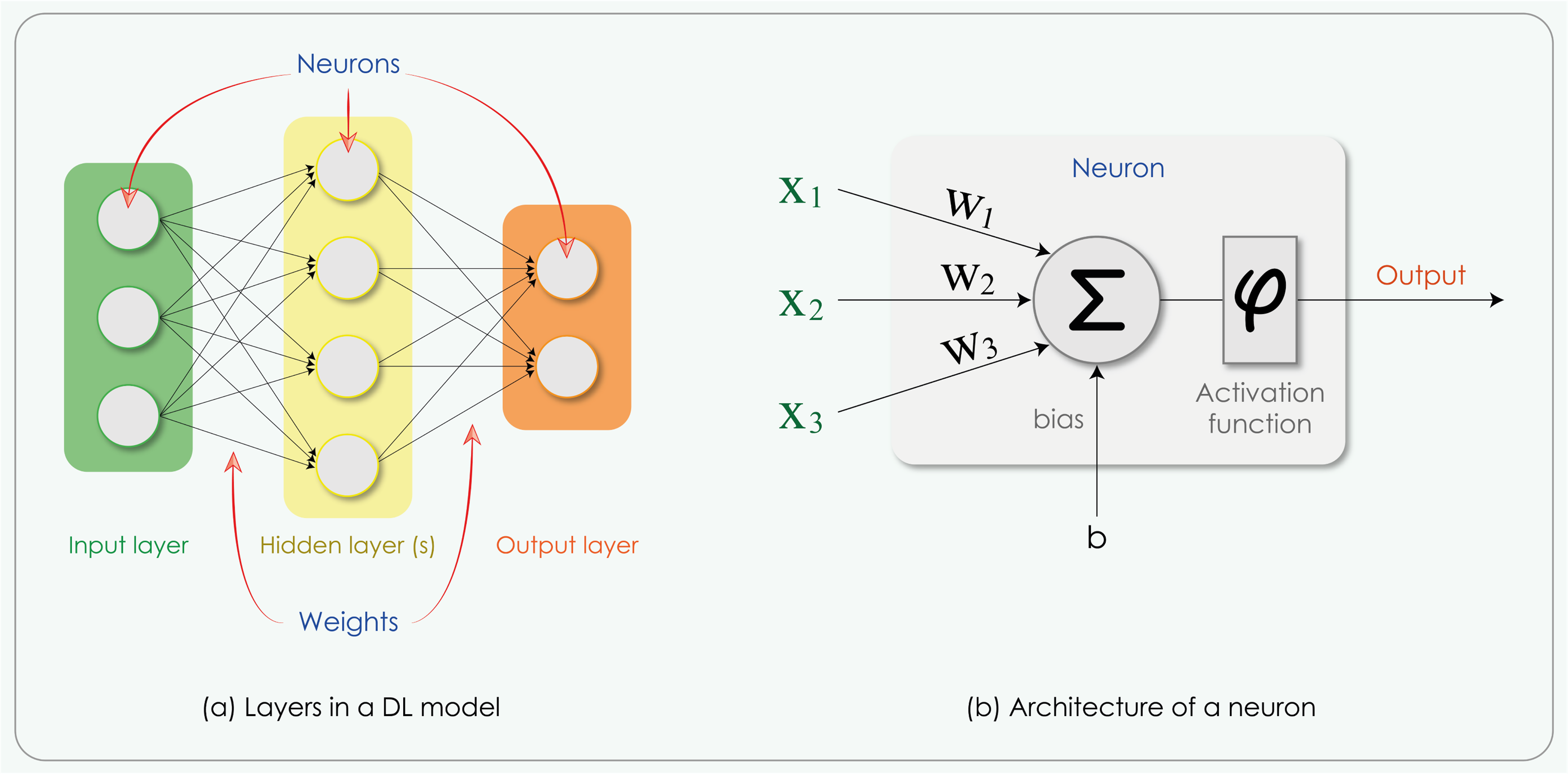}
\caption{Typical composition of DL model.}
\label{img:tcm}
\end{figure} 

\subsection{Architectures}
\begin{figure}[h]
\centering
\captionsetup{justification=centering}
\includegraphics[width=8cm]{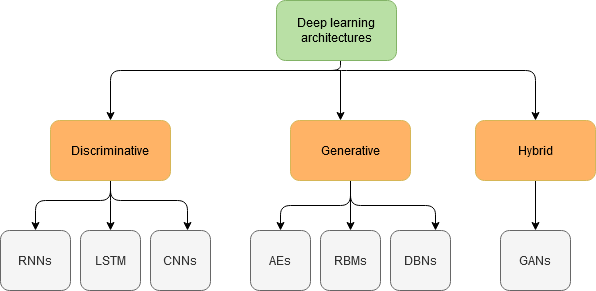}
\caption{Classification schema of DL architectures}
\label{ref:cdl}
\end{figure}

An Artificial Neural Network (ANN) is a generic term to encompass any structure of interconnected neurons which send information to each other. Deep-learning networks (a.k.a. Deep Neural Networks-DNNs) are distinguished from the more commonplace single-hidden-layer neural networks by their depth; that is, the high number of node layers through which data passes in a multistep process of pattern recognition. Note that a network with just one hidden layer cannot actually be considered deep~\cite{IEEEhowto:18}. Fig.~\ref{img:tcm} displays the typical composition of a deep learning model. The main goal of using a deep model is to arrive at the point of least error as fast as possible. Broadly, DL models can be categorized into three categories, namely generative, discriminative, and hybrid models. Discriminative models typically provide supervised learning approaches, while generative models are employed for unsupervised learning. Hybrid models incorporate the benefits of both generative and discriminative models. Fig.~\ref{ref:cdl} depicts the classification schema of DL architectures.

\subsubsection{Deep Convolutional Neural Networks (CNNs)}
In DNN, the layers are fully connected (i.e., all neurons between adjacent layers are connected), which can be a problem if the dimensionality of the input space is enormous. For instance, to model a color image of size $200\times200$ pixels (200 wide, 200 high, three color channels), a single fully-connected neuron in the first hidden layer of a conventional NN would have $200\times200\times3=120000$ weights (parameters). This neuron would be very difficult to train even with a shallow architecture with just one layer. The full connectivity is wasteful, and the enormous number of the involved parameters can  lead to overfitting. CNNs, on the other hand, scale well to full images because they are destined to process data that come in the form of multiple arrays, for example, a color image comprised of three 2D arrays including pixel intensities of the three color channels (RGB). Four fundamental ideas behind CNNs leverage the properties of natural signals, namely local connections, shared weights, pooling, and the use of many layers \cite{IEEEhowto:24}.

Typically, the architecture of CNN is structured in a series of steps. The first layers of a CNN consist of pairs of convolutional (CLs) and pooling layers (PLs), which form a powerful feature extractor block. Subsequently, the convolutional layer applies a specified number of convolution filters (set of weights) to the input data (e.g., an RGB image). These filters perform a convolutional process to transform the input data or the previous layer feature maps into the output feature maps. This process is usually followed by the application of an activation function to the output, such as ReLU, in order to introduce nonlinearities into the model. It is common in CNN architecture to periodically insert a PL in-between successive CLs. Its function is to progressively reduce the dimensionality of the feature map in order to decrease the number of parameters and, consequently, the computation complexity in the network, thus controlling over-fitting. A commonly used pooling algorithm is max pooling, which extracts sub-regions of the feature map and keeps their maximum value while discarding the rest. At the end of the network, one or more fully-connected layers are usually added. CNN has been found highly useful and has been commonly used in computer vision and image recognition. Therefore, using CNNs for image conversion, feature learning in intrusion detection is practicable.
\subsubsection{Recurrent Neural Networks (RNNs)}
The RNN model architecture is a feedback loop linking layer to layer with the ability to store data of previous input, increasing in that way the reliability of the model~\cite{IEEEhowto:25}. The “depth” of an RNN can be as large as the length of the input data sequence. RNNs are very powerful for modeling sequence data. One main problem of RNNs is their sensitivity to the vanishing and exploding gradients~\cite{IEEEhowto:27}. The Long Short-Term Memory (LSTM) \cite{IEEEhowto:26} architecture is used to handle this problem by providing memory blocks in its recurrent connections. Each memory block comprises memory cells, which store the network temporal states. Besides, it contains gated units for controlling the information flow. Given that RNN and its variants are efficient in handling sequential data (e.g., time series data), they can contribute in developing cyber-defense systems for IoT scenarios.  

\subsubsection{Deep Belief Networks (DBNs)}
A DBN~\cite{IEEEhowto:29} is a deep generative model composed of a visible layer and multiple hidden layers of latent variables. There are connections between the layers, but not between units within each layer. DBNs could be used as a feature extraction method for dimensionality reduction. On the other hand, when associating class labels with feature vectors, a DBN is used for classification~\cite{IEEEhowto:19}. Therefore, DBNs not only learn high dimensional representations, but also perform classification tasks in an efficient manner~\cite{IEEEhowto:10}. An unsupervised greedy learning algorithm can be used to pre-train and fine-tune a DBN in order to learn a similarity representation over the nonlinear, high-dimensional input data, something that largely facilitates the classification tasks. Thus, the output of the intrusion detection model could be improved when using a DBN.

\subsubsection{Deep Boltzmann Machine (DBM)}
When a DBM is trained with a large supply of unlabeled data and fine-tuned with labeled data, it acts as a good classifier. Its structure is a derivative of a general Boltzmann machine (BM), which is a network of units based on stochastic decisions to determine their on and off states. Although the BM algorithm is simple to train, it turns out to be slow in the process. A reduction in the number of hidden layers of a DBM to one forms a Restricted Boltzmann Machine (RBM). According to Salakhutdinov~\cite{IEEEhowto:20}, Deep Boltzmann machines are interesting for several reasons. First, like DBNs, DBMs have the potential of learning internal representations that become increasingly complex. Second, high-level representations can be built from a large supply of unlabeled sensory inputs and very limited labeled data can then be used to only slightly fine-tune the model for a specific task at hand. Finally, unlike deep belief networks, the approximate inference procedure in addition to an initial bottom-up pass can incorporate top-down feedback, allowing deep Boltzmann machines to better propagate uncertainty, and hence deal more robustly with ambiguous inputs.

\subsubsection{Auto-Encoders (AEs)}
Auto-Encoders~\cite{IEEEhowto:30} are neural networks that are trained to reconstruct their input using an intermediate representation (code). Auto-Encoders and RBMs are very similar models because they share the same architecture: an input layer that represents the data, a hidden layer that represents the code to be learnt, and the weighted connections between them. However, The Auto-Encoder is trained by minimizing the reconstruction error, so an extra layer is added on the top to represent a reconstruction of the original data. Both sets of weights are tight, which means that they are actually the same weights. These models also have the capability of extracting good representations from unlabeled data that work well to initialize deeper networks~\cite{IEEEhowto:18}. One of the main problems of the Auto-Encoder model is that it could potentially learn a useless identity transformation when the representation size (the hidden layer) is larger than the input layer (the so-called over-complete case). To overcome this potential limitation, there is a simple variant of the model called Denoising Auto-Encoders (DAE)~\cite{IEEEhowto:21} that works well in practice. The basic idea is to feed the encoder/decoder system with a corrupted version of the original input, in order to force the system to reconstruct the clean version. This small change allows the DAE to learn useful representations, even in the over-complete case. In practice, the AE architectures can learn significant higher level representation (features) from raw traffic data, which is a remarkable capability for the intrusion detection task.

\subsubsection{Generative Adversarial Networks (GANs)} GANs~\cite{IEEEhowto:31} are comprised by two main components, namely the generative and discriminator networks (i.e., the generator and the discriminator). The generator is responsible for generating new data after learning the distribution from a training set of real data. The discriminator is responsible for distinguishing the real from the fake data that have been generated by the generator. Applications where GAN is often deployed include image generation, image transformation, image synthesis, and image super-resolution. GANs can be employed for scenarios where the creation of something new out of the available data is required. Furthermore, this architecture can be used to generate adversarial attacks or malware samples to deceive the cyber-defense systems~\cite{IEEEhowto:36}. 
\section{DL framework for cybersecurity applications}
\label{sec:framework}
\begin{figure}[h]
\centering
\captionsetup{justification=centering}
\includegraphics[width=8.5cm]{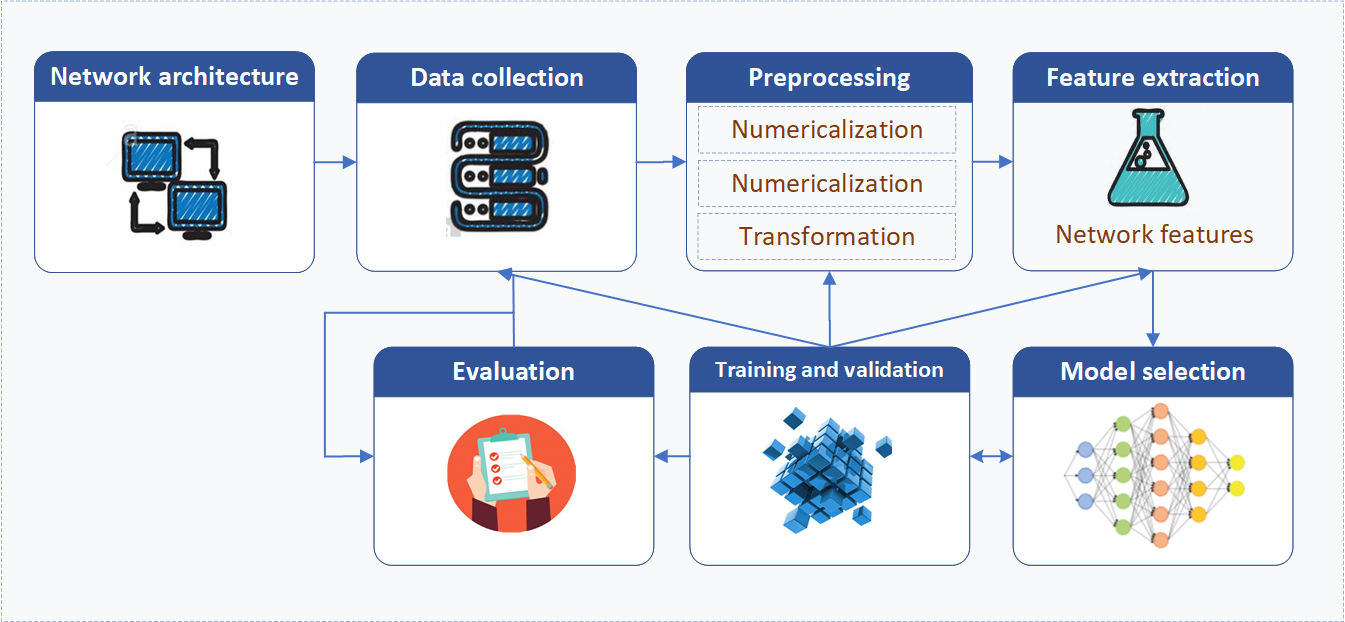}
\caption{DL framework for cybersecurity applications.}
\label{img:df}
\end{figure}

Herein, we introduce a DL framework for Network Intrusion Detection. The conceptual model of the proposed framework is shown in Fig.~\ref{img:df}. First, a sufficient volume of data is collected from the network infrastructure. Access to a large amount of data makes it possible to solve challenging and complex security problems. Sufficient data quality and quantity are fundamental. Consider that whenever we collect data, we need to collect it in such a way as to learn general trends. Then, in the preprocessing step, the collected data should be cleaned, merged and converted to structural formats, while data sources are mapped to each other. The feature engineering step refers to the extraction and selection of useful features, which are not only critical to defining and enriching the predictors, but also play a vital role in machine learning. However, in practice, coming up with appropriate features is often challenging, labor-intensive, time-consuming, and requires expert knowledge and ingenuity. An alternative for discovering such features or representations is representation learning, which is based on recognizing and disengaging the latent explanatory factors that exist in the data~\cite{IEEEhowto:19}. In that way, through learning representations of the data, the extraction of useful information when constructing classifiers or other predictors is facilitated.

The choice of the DL model is highly dependent on the input features, which directly affect the accuracy of the model. The modelling process is iterative, as it gives crucial insights related to the refinement of data preparation and model specification. It is significantly helpful to try several algorithms with specific parameters (and hyper-parameters) to find the ideal model. Typically, a dataset is divided into three parts, namely a training, a validation, and a test set. The first set is used for training the model, whose prediction accuracy is measured on the validation set. The validation set accuracy is one of the principal criteria in accepting or rejecting the trained model. When we are satisfied with the selected model type and hyper parameters, the next step should be to train a new model with the entire set of available data using the best hyper-parameters found. This should include any data that was previously held aside for validation. The last step is periodic evaluation over updated test sets, which is tremendously useful to verify that the model can detect and predict zero days attacks.

\section{Literature review}
\label{sec:review}
\begin{figure}[h]
\centering
\captionsetup{justification=centering}
\includegraphics[width=6.5cm]{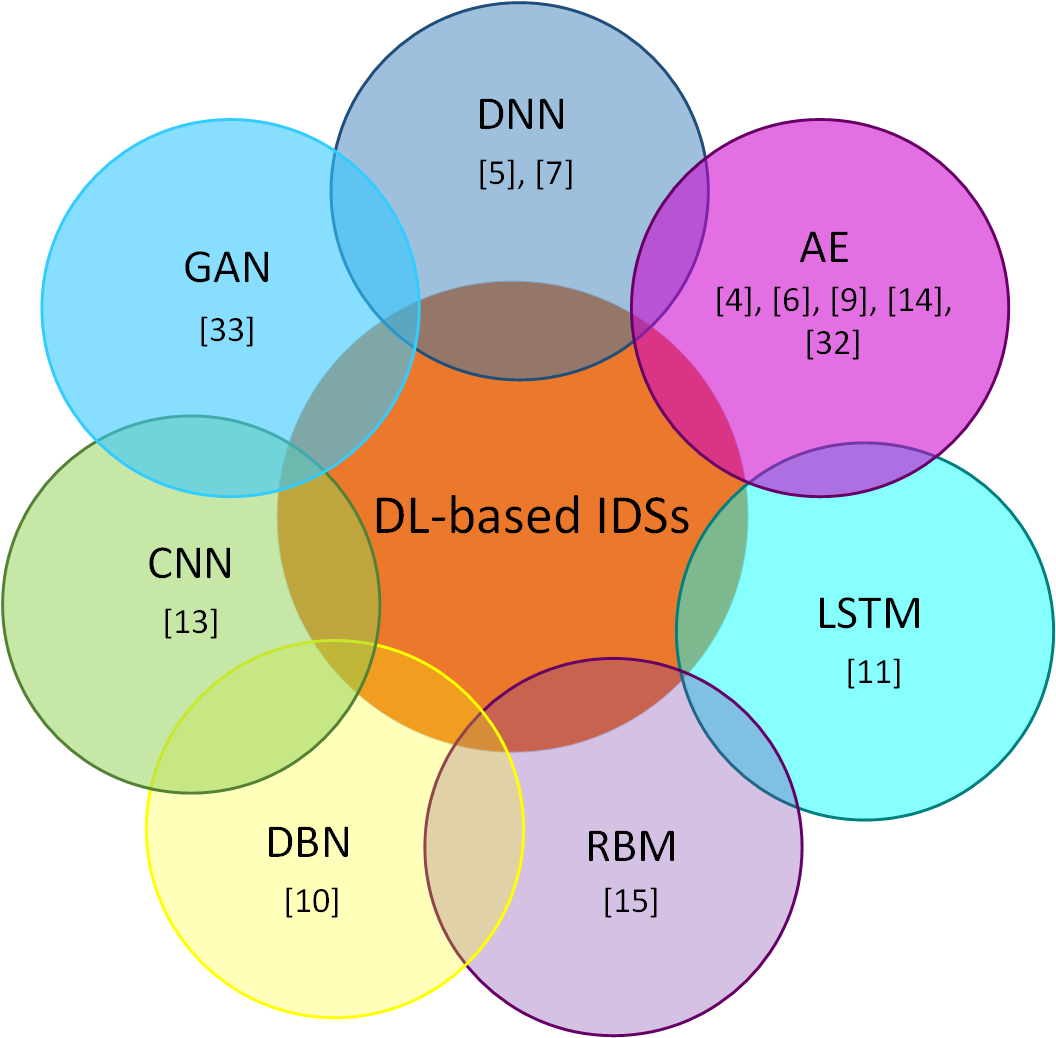}
\caption{Classification of selected DL-based IDSs}
\label{img:cdlids}
\end{figure}

This section focuses mainly on providing information about intrusion detection systems leveraging deep learning models. Fig.~\ref{img:cdlids} illustrates the classification of some of the examined intrusion detection methods based on the deep learning model they employ. It can be used for a quick overview of positioning of each included approach in the state-of-the-art.

In the early 2000s, several studies presented two issues in developing an effective and flexible Network Intrusion Detection System (NIDS) for unknown future attacks. The first issue was that it was difficult to extract features from the network traffic dataset for anomaly detection. The second issue was that a labeled (trained) traffic dataset from real network traffic was not available. In order to overcome these, Fiore et al.~\cite{IEEEhowto:15} used a Restricted Boltzmann Machine to implement semi-supervised anomaly detection systems where the classifier was trained with "normal" traffic data only, so that knowledge about anomalous behaviors was constructed and evolved in a dynamic way. In the experimental phase, the accuracy of RBM was tested in classifying normal data and data infected by botnet attacks. A second experiment trained RBM with the KDDCup 99 dataset and tested it against real world data. To randomize the order of test data, the experiment was repeated 10 times. 
The results revealed that when the classifier was tested in a network widely different from the one where training data were taken from, performance decayed. This suggests the need for further investigation over the nature of anomalous traffic and the intrinsic differences with normal traffic. 

Meanwhile, Li et al.~\cite{IEEEhowto:4} demonstrated that the use of the AE deep learning method is effective for achieving data dimension reduction and it can improve the detection accuracy. In their experimental model, the authors applied the AE in order to convert complicated high-dimensional data into low dimensional codes with a nonlinear mapping, thereby reducing the dimensionality of data, extracting its main features, and then applying the DBN learning method to detect malicious code. The comparative analysis between single DBN and the DBN + AE yielded accuracy rates of $89.75 \%$ and $91.4 \%$, respectively. Consequently, the hybrid malicious code detection model was superior to that of the single DBN. 

Javaid et al.~\cite{IEEEhowto:6} applied deep learning techniques such as sparse AE and SMR. Based on those techniques, this research introduced two main steps in terms of feature extraction and supervised classification. The first step was to collect unlabeled network traffic data. Then, the next step was to apply the extracted features to a labeled traffic dataset, which could be collected in a confined, isolated, and private network environment for supervised classification. The primary goal of this research was to evaluate the performance of deep learning based on accuracy. To evaluate the classification accuracy, this research employed the NSL-KDD training dataset using $\textit{2-class}$ (normal and attack), $\textit{5-class}$ (normal and four different attack categories), and $\textit{23-class}$ (normal and twenty-two different attacks). According to the results, DL showed better accuracy performance for $\textit{2-class}$ compared to SMR; however, there was no significant improvement for $\textit{5-class}$ and $\textit{23-class}$. The accuracy for $\textit{2-class}$ was $88.39 \%$, which was much higher than SMR $(78.06\%)$. It should be noted that the highest accuracy achieved using the NB-Tree methodology was $82\%$.

Ashfaq et al.~\cite{IEEEhowto:7} improved the classifier's performance for IDS through a novel fuzziness based semi-supervised learning approach, utilizing unlabeled samples assisted with a supervised learning algorithm. A single hidden layer feed-forward neural network (SLFN) was trained to output a fuzzy membership vector, and the sample categorization (low, mid, and high fuzziness categories) on the unlabeled samples was performed using the fuzzy quantity. The classifier was trained after incorporating each category separately into the original training set. During the experiment phase, the authors performed necessary scaling to normalize the data. According to the results, the accuracy obtained by the proposed algorithm was the highest compared to those obtained with J48, Naive Bayes, NB tree, Random forests, Random tree, Multi-layer perceptron, and Support Vector Machine (SVM). 

Diro et al.~\cite{IEEEhowto:5} introduced a self-taught DL scheme in which unsupervised feature learning has been applied on the NSL-KDD dataset. It utilizes a novel model of parallel training and parameter sharing by local fog nodes and it detects network attacks in the distributed fog-to-things networks following a deep learning approach. The outputs of the model training on the distributed fog nodes are the attack detection models and their associated local learning parameters. These local parameters are sent to the coordinating fog node for global update and re-propagation. This sharing scheme results in better learning, as it enables the sharing of the best parameters avoiding in this way local overfitting. The results of the experiments give rise to two conclusions. First, the distributed model has a better performance than the centralized model, since the increased number of nodes in the distributed network of Fog systems, leads to a $3\%$ increase in the overall accuracy of detection from around $96\%$ to over $99\%$. Second, the detection rate shows that deep learning is better than classic machine learning for both binary and multi-classes. 

Yu et al.~\cite{IEEEhowto:9} demonstrated the application of unsupervised DL techniques to automatically learn essential features from raw network traffic counts. The authors implemented an SDA architecture to detect traffic generated by botnets. The procedure of SDA was split into two stages. The first stage, unsupervised layer-wise pre-training, is a greedy layer-wise training process. In the second stage (i.e., supervised fine-tuning stage), a logistic regression layer for classification was added on top of the stacked denoising autoencoders.  In the experiment phase, the UNB ISCX IDS 2012 dataset was used \cite{IEEEhowto:32}. Two datasets of different sizes were constructed. One consisting of $43\%$ of the UNB ISCX IDS 2012 and another using the original dataset in its entirety. The experiment was divided into three parts: binary classification using the SDA-based deep neural network, multi-classification using the SDA-based deep neural network, and classification using different DL architectures. According to the results, SDA achieved better overall performance in almost all experiments except multi-classification on the $43\%$ dataset. The authors demonstrated that SDA employing the denoising criterion can learn significant higher level representation (features) from raw traffic data, and thus they concluded that DL approaches have remarkable capabilities for the intrusion detection task.

Yin et al.~\cite{IEEEhowto:11} proposed an IDS based on RNNs. The principal goal of their study was to demonstrate that RNN is suitable for the development of a classification model with higher accuracy and performance than the traditional machine learning classification methods in binary and multiclass classification. To evaluate the classification accuracy, this research used the NSL-KDD dataset in the experimental phase. The experiment consisted of two parts: the study of the performance of the RNN-IDS model for binary classification (Normal, anomaly) and five category classification (Normal, DoS, R2L, U2R and Probe). In the binary classification and the five-category classification, they compared the performance of the RNN model with that of an ANN, naive Bayesian, random forest, support vector machine, random tree and multilayer perceptron. The results in the binary classification showed that the RNN model achieved a detection rate of $83.28\%$ when given $100$ epochs. Meanwhile, the algorithms J48, Naive Bayesian, Random Forest, Multi-layer Perceptron, Support Vector Machine and ANN obtained a detection rate of $81.2\%$.  In the five-category classification the RRN model obtained an accuracy rate of $81.29\%$, which was better than the $79.9\%$ obtained by using J48, Naive Bayesian, Random Forest, Multi-layer Perceptron, Support Vector Machine and ANN. In order to increment the accuracy rate in the RNN model, they applied the reduced-size technique on the dataset, reaching in that way up to $97.09\%$ (a gain of about $17.19\%$).  

Li et al.~\cite{IEEEhowto:13} employed a visual conversion of the NSL-KDD dataset format to evaluate the performance of CNN in detecting novel attacks. In the experimental phase, the data were categorized into five classes: Normal, Dos, Probe, U2R, and R2L. The NSL-KDD feature attributes were classified into three groups: basic features, traffic features, and content features. Each sample of the NSL-KDD dataset contained 41 features (integer, float, symbolic, or binary). To convert NSL-KDD data format into a visual image type, they mapped various types of features into binary vector space and then transformed the binary vector into the image. After that, the NSL-KDD data form turned into a binary vector with 464 dimensions. Then, they turned every $8$ bits into a grayscale pixel. The binary vector with $464$ dimensions was transformed into an $\mathbf{8\times8}$ grayscale image with vacant pixels padded with $0$. The NSL-KDD dataset was divided into training and test sets. In order to test intrusion detection and identify the ability to discover new attacks, $17$ additional attack types were incorporated in the test set. The results showed that the proposed method had good performance on the NSL-KDD dataset in comparison to J48, Naive Bayes, NB Tree, Random forest, Random tree, Multi-layer perceptron, and SVM. The CNN model obtained $81.57\%$ of accuracy rate, whereas, the other algorithms achieved accuracy rates of $63.97\%$, $55.77\%$, $66.16\%$, $62.26\%$, $58.51\%$, $57.34\%$, $42.29\%$ and $81.57\%$ respectively.

Shone et al~\cite{IEEEhowto:14} proposed a non-symmetric deep auto-encoder (NDAE) for unsupervised feature learning. The classifier model was constructed by stacked NDAEs on the KDD Cup'99 and NSL-KDD datasets. Fundamentally, this involves the proposed shift from the encoder-decoder paradigm (symmetric) towards utilizing just the encoder phase (non-symmetric).  The reasoning behind this is that given the correct learning structure, it is possible to reduce both computational and time overheads, with minimal impact on accuracy and efficiency. NDAE was used as a hierarchical unsupervised feature extractor. The model was realized by stacking NDAEs in order to create a DL hierarchy. Stacking the NDAEs offered a layer-wise unsupervised representation learning algorithm, which allowed the model to learn the complex relationships between different features. It also had feature extraction capabilities. Hence, it was able to refine the model by prioritizing the most descriptive features. They combined the DL power of stacked NDAEs with a shallow learning classifier. Random Forest (RF) was used as shallow learning classifier in order to increase the classification power of the stacked autoencoders. The model trained the RF classifier using the encoded representations learned by the stacked NDAEs to classify network traffic into normal data and known attacks. In the evaluation phase, they compared the stacked NDAE model against the mainstream DBN technique. These comparisons have demonstrated that the model offers up to a $5 \%$ improvement in accuracy and up to $98.81 \%$ of training time reduction.

Today, Internet of Things (IoT) devices are becoming more prevalent and by their very nature produce an enormous amount of multi/univariate data~\cite{IEEEhowto:33}. At the same time, due to IoT devices being connected to the Internet, they are vulnerable to various attacks. In such scenarios, a large amount of unlabeled or semi-labeled data is produced. DL provides different techniques, such as AEs and GANs, for handling unlabeled data and learn useful patterns in an unsupervised (more human-like) fashion efficiently. For instance, in~\cite{IEEEhowto:33}~\footnote{The source code of this study is publicly available at https://github.com/MayraMacasC/AnomalyDetection}, a deep autoenoder is used for performing unsupervised anomaly detection in order to automatically discover intrusions and anomalies in critical infraestructures. This study demonstrated high accurancy rate and fast time for convergence.  In order to generate fake network data, Yin et al. \cite{IEEEhowto:34} employed GAN with the aim of improving the  performance of an original botnet detection model. In the architecture of GAN, a 4-layer DNN was used as the discriminator and a 3-layer LSTM was adopted as the generator. They used 16 features based on network flow and conducted several experiments on the ISCX botnet dataset \cite{IEEEhowto:35}. Their model decreased the false positive rate from 19.19\% to 15.59\%. 

Another way for dealing with the huge amount of unlabeled data in the cybersecurity domain is applying DL methods in a semi-supervised fashion. For example, Gao et al.~\cite{IEEEhowto:10} built an intrusion detection model based on DBN, where the DBN was pre-trained in an unsupervised greedy layer-wise manner for learning a stack of RBMs by the Contrastive Divergence (CD) algorithm. The output feature representation of each RBM was employed as the input data for training the next RBM in the stack. Finally, after the pre-training, the DBN was fine-tuned in a supervised manner employing backpropagation (BP) of error derivatives, and the initial weights and biases of each layer were corrected. In the experiment phase, the KDD Cup 1999 dataset was used and the resuls showed that the performance of the DBN model using an RBM network with three or more layers outperformed SVM and ANN. 
\section{Conclusion}
\label{sec:Conclusion}
Motivated by the challenges in the cybersecurity domain and the current state-of-the-art for DL applications in a multitude of areas, in this work we have surveyed and classified the DL techniques employed for intrusion detection.
The results of the examined models demonstrate that the application of DL in cybersecurity and in particular in IDS is promising. In spite of the significant advancements, however, there is still much room for future research on a variety of open issues, such as new approaches to testing the reliability and efficiency of knowledge-based and behavioral approaches in intrusion detection. Generating adversarial examples to confuse DL-based IDSs and malware detection systems needs to be further investigated.

\ifCLASSOPTIONcaptionsoff
  \newpage
\fi

\end{document}